\documentclass[aps,float,prd,psfig,amsmath,amssymb,12pt]{revtex4}

\usepackage[dvipdfmx]{graphicx}
\usepackage{dcolumn}
\usepackage{bm}
\usepackage{braket}
\usepackage{breqn}
\usepackage{color}
\usepackage{hyperref}

\begin{document}
\begin{titlepage}
\unitlength = 1mm
\begin{flushright}
KOBE-COSMO-23-07
\end{flushright}

\vskip 1cm
\begin{center}

{ \large \textbf{
Hellings-Downs curve deformed by ultralight vector dark matter
}}
\vspace{1.8cm}
\\
Hidetoshi Omiya$^{\flat}$, Kimihiro Nomura$^{\flat}$, Jiro Soda$^{\flat,\sharp}$
\vspace{1cm}

\shortstack[l]
{
\it $^\flat$ Department of Physics, Kobe University, Kobe 657-8501, Japan\\
\it $^\sharp$ International Center for Quantum-field Measurement Systems for Studies
of the Universe \\ 
\it \ \  and Particles (QUP), KEK, Tsukuba 305-0801, Japan
}

\vskip 4.0cm

{\large Abstract}\\
\end{center}

Pulsar timing arrays (PTAs) provide a way to detect gravitational waves (GWs) at nanohertz frequencies. To ensure the detection of GWs,  observational data must exhibit the Hellings-Downs angular correlation. 
It is also known that PTAs can probe ultralight dark matter. In this paper, we consider possible contamination of the Hellings-Downs angular correlation by the ultralight dark matter. We find that ultralight vector dark matter can give rise to the deformation of the Hellings-Downs correlation curve. Thus, the Hellings-Downs correlation curve could contain information on ultralight dark matter with a spin.
\vspace{1.0cm}
\end{titlepage}

\hrule height 0.075mm depth 0.075mm width 165mm
\tableofcontents
\vspace{1.0cm}
\hrule height 0.075mm depth 0.075mm width 165mm

\section{Introduction}
The first direct detection of gravitational waves (GWs) by LIGO~\cite{LIGOScientific:2016aoc} has initiated GW astronomy. To further develop GW astronomy,
observing GWs in a different frequency range is crucial. Intriguingly, pulsar timing arrays (PTAs) provide a way to detect gravitational waves (GWs) at nanohertz frequencies~\cite{Detweiler:1979wn}.
A requirement for claiming the detection of GWs~\cite{Allen:2023kib} is to verify Hellings-Downs angular correlation~\cite{Hellings:1983fr}. 
Recently, evidence of stochastic GWs has been reported~\cite{NANOGrav:2023gor, NANOGrav:2023hvm, Antoniadis:2023ott, Xu:2023wog, Reardon:2023gzh}.
In particular, the angular correlation seems to follow the Hellings-Downs pattern.

Curiously, PTAs can also explore the ultralight dark matter. Ultralight dark matter is a model of dark matter realized by the coherent oscillation of an ultralight bosonic field, with its frequency given by the mass of the field. This coherent oscillation will then lead to fluctuations of the gravitational potential, which cause the timing residual of the pulses from the pulsar~\cite{Khmelnitsky:2013lxt}. Since PTAs are sensitive to the nanohertz frequency band, the mass around $10^{-24}~{\rm eV}\lesssim \mu \lesssim 10^{-22}~{\rm eV}$ can be explored.
Indeed, the energy density of axion dark matter has been constrained by pulsar timing observations~\cite{Porayko:2014rfa, Porayko:2018sfa, Kato:2019bqz}. 

Although the axion dark matter is more popular for its simplicity, there is a possibility of ultralight vector dark matter~\cite{Nelson:2011sf, Nakayama:2019rhg}. The interesting feature of vector dark matter is introducing a particular direction determined by the local direction of the vector field.
Remarkably, it turns out that signals of ultralight vector dark matter in PTAs are strongly anisotropic due to this feature~\cite{Nomura:2019cvc}. 

Observationally, it is true that GWs and dark matter coexist. Hence,  it is important to clarify the effect of ultralight dark matter on the Hellings-Downs correlation. However, the axion dark matter cannot produce a feature in angular correlation due to its scalar nature.
On the other hand, the vector dark matter has the potential to induce a nontrivial angular correlation pattern since a coherently oscillating vector field defines a specific direction.
Thus, the Hellings-Downs curve might be contaminated by the vector dark matter.

For this purpose, we examine the angular correlation of PTAs
in the coexistence of GWs and vector dark matter. By calculating the cross-correlation signal from the pulsars, we find that the vector dark matter induces the mixture of the monopole and the quadrupole angular correlation pattern. Therefore, the Hellings-Downs correlation is deformed by the vector dark matter. It is expected that the same happens for any dark matter with spin. Hence, the Hellings-Downs correlation tells us not only the existence of GWs but also that of ultralight dark matter.

The organization of the paper is as follows.
In section II, we evaluate pulsar timing residual induced by ultralight vector dark matter. In section III, we calculate the angular correlation of pulsar timing signals and show that the Hellings-Downs curve will be deformed by the vector dark matter. In section IV, we summarize this paper. We fix our units to $c = \hbar = 1$ in the following.

\section{Pulsar timing residual induced by vector dark matter}

In this paper, we focus on the ultralight vector dark matter denoted by $A_\mu$. The coherence length of the vector field is estimated to be
\begin{align}
	\lambda_{\rm dB} = \frac{2\pi}{\mu v_{\rm DM}} \sim 0.4 \,{\rm kpc} \left(\frac{10^{-22} {\rm eV}}{\mu}\right) \left(\frac{10^{-3}}{v_{\rm DM}}\right)~,
\end{align}
where $\mu$ is the mass of the ultralight field and $v_{\rm DM}$ is the typical velocity of the dark matter particle in the unit of the speed of light. 
Another important number is the occupation number of the dark matter particle, which is roughly calculated to be
\begin{align}
	N_{\rm DM} \sim \frac{\rho_{\rm DM}}{\mu}\lambda_{\rm dB}^3 \sim 10^{93}\left(\frac{\rho_{\rm DM}}{0.4 {\rm GeV\cdot cm^{-3}}}\right) \left(\frac{10^{-22} {\rm eV}}{\mu}\right)^4 \left(\frac{10^{-3}}{v_{\rm DM}}\right)^3.
\end{align}
This huge occupation number allows us to treat the ultralight field as a classical field.

The action of the vector dark matter $A_\mu$ is given by
\begin{align}
	S &= \int d^4 x \sqrt{-g}\ \left(-\frac{1}{4}F_{\mu\nu}F^{\mu\nu} - \frac{1}{2}\mu^2 A_\mu A^\mu\right)~,
\end{align}
with
\begin{align}
	F_{\mu\nu} &= \nabla_\mu A_\nu - \nabla_\nu A_\mu~.
\end{align}
Taking variation of the action, we obtain the equation of motion for $A_\mu$ as
\begin{align}\label{eq:eom}
	\nabla_\mu F^{\mu\nu} - \mu^2 A^\nu = 0~.
\end{align}
Since $F_{\mu\nu}$ is anti-symmetric, the divergence of Eq.~\eqref{eq:eom} leads to
\begin{align}
	\nabla_\mu A^\mu =0 ~,
\end{align}
which immediately gives
\begin{align}
	A^0 \sim \frac{1}{\mu \lambda_{\rm dB}} A^i~.
\end{align}
Here we utilized the fact that the typical scale of the spatial variation of the field is $\sim \lambda_{\rm dB}$, and the vector field is coherently oscillating with frequency $\mu$. Since $\mu \lambda_{\rm dB} = 2\pi/v_{\rm DM} \gg 1$, we can neglect the $0$-th component of the vector field.

To summarize, we can approximate the vector field configuration to be
\begin{align}
	A_0 &= 0~, & \bm{A}(t,\bm{x}) &= \bm{\Omega}_A(\bm{x})  A(\bm{x}) \cos\left(\mu t + \alpha(\bm{x})\right)~,
\end{align}
where  $\bm{\Omega}_A(\bm{x})$ is a unit vector pointing to the direction of the vector field and  $\alpha(\bm{x})$ describes the spatial dependence of the field. 
Note that the spatial gradient of the field is on the scale of $2\pi / \lambda_{\rm dB} = \mu v_{\rm DM}$, which is much smaller than $\mu$.
In the following, we assume
\begin{align}
	\alpha(\bm x) &= \bm{k}\cdot \bm{x}~
\end{align}
with $|\bm{k}| \ll \mu$. Here, we treated the field as completely monochromatic since the expected width of the frequency should be much smaller than the frequency resolution of pulsar timing observations~\cite{NANOGrav:2023gor, NANOGrav:2023hvm, Antoniadis:2023ott, Xu:2023wog, Reardon:2023gzh}.
The amplitude $A(\bm{x})$ is related to the dark matter density as
\begin{align}
	\rho_{\rm DM} &\sim \frac{\mu^2}{2}A^2(\bm{x})~.
\end{align}

The coherent oscillation of the vector dark matter produces the fluctuations of the metric.
Solving Einstein equations in the weak field limit, we obtain~\footnote{It is not obvious that we can bring the metric perturbation to the form~\eqref{eq:s1metricpert}. However, under the assumption of neglecting the spatial gradient, we can show that it is true. See~\cite{Nomura:2019cvc} for the derivation.}
\begin{align}\label{eq:s1metricpert}
	ds^2 = -(1-2\Psi(t,\bm{x})) dt^2 + \left[(1+ 2\Psi(t,\bm{x}))\delta_{ij} + \gamma_{ij}(t,\bm{x})\right]dx^i dx^j~,
\end{align}
with
\begin{align}
	\Psi(t,\bm{x}) &= \Psi_{\rm osc}(\bm{x})\cos\left(2\mu t + 2 \bm{k}\cdot \bm{x}\right)~,\\
	\gamma_{ij}(t,\bm{x}) &= h_{{\rm osc}}(\bm{x}) \left(\delta_{ij} - 3 \Omega_{A,i}(\bm{x})\Omega_{A,j}(\bm{x}) \right)\cos(2\mu t + 2 \bm{k}\cdot \bm{x})~,\\
	\Psi_{\rm osc}(\bm{x}) &= -\frac{\pi G }{6} A^2(\bm{
x})~,\\
	h_{{\rm osc}}(\bm{x}) &= \frac{4 \pi G }{3}A^2(\bm{
x}) = -8 \Psi_{\rm osc}(\bm{x})~.
\end{align}
Here we only considered the time-dependent parts, which affect the observed periodicity of pulses from pulsars. The details of deriving Eq.~\eqref{eq:s1metricpert} can be found in~\cite{Nomura:2019cvc}.

The metric fluctuations induce a redshift of the observed pulses from pulsars.
The observed redshift of pulses at time $t$ from the pulsar $a$ is defined by 
\begin{align}
    z_a(t) = \frac{\omega_{0,a} - \omega_{\text{obs},a}(t)}{\omega_{0,a}} ~,
\end{align}
where $\omega_{0, a}$ is the intrinsic angular frequency of pulses emitted from the pulsar $a$, and $\omega_{\text{obs}, a}(t)$ is the observed angular frequency of pulses which are affected by metric perturbations.
Due to the metric perturbations~\eqref{eq:s1metricpert}, the redshift is given by~\cite{maggiore2018gravitational}
\begin{align}\label{eq:redshift}
    z_a(t) 
    &= \int_{t - L_a}^t dt' \, \bigg[ \frac{\partial}{\partial t'} \Psi(t', \bm{x}) \bigg]_{\bm{x} = \bm{x}_a(t')} 
    + \frac{1}{2} u_a^i u_a^j \int_{t - L_a}^t dt' \, \bigg[ \frac{\partial}{\partial t'} \gamma_{ij}(t', \bm{x}) \bigg]_{\bm{x} = \bm{x}_a(t')} 
\end{align}
where $\bm{x}_a(t') = (t-t') \bm{u}_a + \bm{x}_{\mathrm{E}}$ with $\bm{x}_{\mathrm{E}}$ being the position of the Earth, and $L_a$ is the distance to the pulsar $a$ from the Earth.
Here, $\bm{u}_a$  is a unit vector pointing to the pulsar $a$.

\begin{figure}
\centering
\includegraphics[keepaspectratio, scale=0.5]{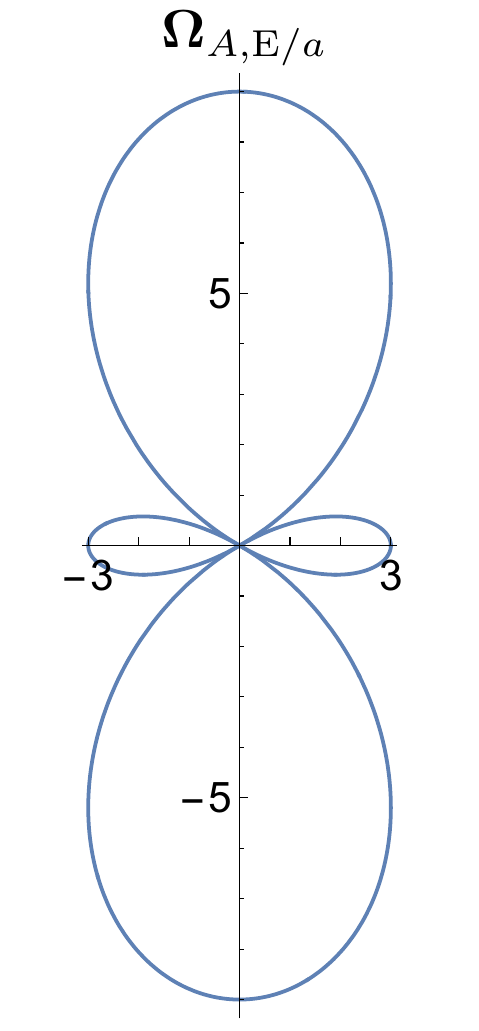}
\caption{The angular dependence of the beam pattern functions $F^{\rm DM}_{a,{\rm E/P}}$ is depicted. The direction of the vector field $\bm{\Omega}_{A,{\rm E}/a}$ is taken to be along the vertical axis, and the angle $\bm{u}_a \cdot \bm{\Omega}_{A,{\rm E}/a}$ is measured from the vertical axis.} 
\label{fig:ang}
\end{figure}

By substituting Eq.~\eqref{eq:s1metricpert} into Eq.~\eqref{eq:redshift}, we obtain~\cite{Nomura:2019cvc}
\begin{align}
    z_a(t) 
    &\approx 
     \Psi_{\mathrm{osc}}(\bm{x}_{\mathrm{E}}) 
    \cos (2\mu t + 2\bm{k}\cdot \bm{x}_{\rm E} )
    - 
    \Psi_{\mathrm{osc}}(\bm{x}_{a}) 
    \cos (2\mu t - 2\mu L_a + 2\bm{k}\cdot \bm{x}_{a}  )
    \notag \\
    &\qquad +  
    \frac{1}{2} 
    h_{\mathrm{osc}}(\bm{x}_{\mathrm{E}}) 
    (1 - 3 (\bm{u}_a \cdot \bm{\Omega}_{A, \mathrm{E}})^2 )
    \cos (2\mu t + 2\bm{k}\cdot \bm{x}_{\rm E} )
    \notag \\
    &\qquad\qquad  - 
    \frac{1}{2} 
    h_{\mathrm{osc}}(\bm{x}_{a}) 
    (1 - 3 (\bm{u}_a \cdot \bm{\Omega}_{A, a})^2 )
    \cos (2\mu t - 2\mu L_a + 2\bm{k}\cdot \bm{x}_{a} )\cr
    &= F_{a,{\rm E}}^{{\rm DM}} \Psi_{\rm osc}(\bm{x}_{\rm E}) \cos (2\mu t + 2\bm{k}\cdot \bm{x}_{\rm E} ) 
    - F_{a,{\rm P}}^{{\rm DM}} \Psi_{\rm osc}(\bm{x}_a) \cos (2\mu t - 2\mu L_a + 2\bm{k}\cdot \bm{x}_{a}  )~,
\end{align}
where $\bm{x}_{\rm E}$ is the position of the Earth, $\bm{x}_{a}$ is the position of the pulsar $a$, and 
$\bm{\Omega}_{A,\mathrm{E}} = \bm{\Omega}_{A}(\bm{x}_{\mathrm{E}})$ and $\bm{\Omega}_{A, a} = \bm{\Omega}_{A}(\bm{x}_a)$ are the directions of the vector field at the Earth and the pulsar $a$, respectively. 
Here, we introduced the beam pattern functions in the presence of the vector dark matter as
\begin{align}
    F_{a,{\rm E}}^{{\rm DM}} &= -3\left(1 - 4 (\bm{u}_a \cdot \bm{\Omega}_{A, {\rm E}})^2 \right)~, &
    F_{a,{\rm P}}^{{\rm DM}} &= -3\left(1 - 4 (\bm{u}_a \cdot \bm{\Omega}_{A, a})^2 \right)~.
\end{align}
The beam pattern functions are functions of the angle between the direction to the pulsar $\bm{u}_a$ and the direction of the vector field $\bm{\Omega}_{A,{\rm E}/a}$.
We show the angular dependence of $F^{\rm DM}_{a,{\rm E/P}}$ in Fig.~\ref{fig:ang}. We observe that the beam pattern functions of the dark matter show a quadrupole pattern.

The timing residual $\Delta T_a(t)$ of the pulsar $a$ can be obtained from the redshift as
\begin{align}\label{eq:residual1}
    \Delta T_a(t) = \int_0^t dt' \, z_a(t')~. 
\end{align}
After performing the integral in Eq.~\eqref{eq:residual1}, we obtain 
\begin{align}	\label{eq:residual2}
	\Delta T_a(t) &=  \frac{1}{2 \mu} \left(F_{a,{\rm E}}^{{\rm DM}} \Psi_{\rm osc}(\bm{x}_{\rm E}) \sin (2\mu t + 2\bm{k}\cdot \bm{x}_{\rm E} )
 - F_{a,{\rm P}}^{{\rm DM}} \Psi_{\rm osc}(\bm{x}_a) \sin (2\mu t - 2\mu L_a + 2\bm{k}\cdot \bm{x}_{a} ) \right)~.
\end{align}
Note that the origin of time can be taken arbitrarily.
We see that the ultralight vector field will induce a timing residual oscillating with a frequency $2\mu$. 

\section{Deformation of Hellings-Downs angular correlation}

 The essential ingredient to detect the GW background is taking the cross-correlation between the noise-independent data~\cite{Detweiler:1979wn,1987MNRAS.227..933M, Hellings:1983fr, Flanagan:1993ix}. By taking cross-correlation, the signal is enhanced relative to the noise. The cross-correlation between the timing residuals of the pulsar $a$ and $b$ can be calculated through
\begin{align}\label{eq:defcorr}
	C_{ab}(\tau) \equiv \braket{\Delta T_a(t)\Delta T_b(t + \tau)} - \braket{\Delta T_a(t)}\braket{\Delta T_b(t + \tau)} ~.
\end{align}
Here, $\braket{\cdots}$ means a time average or an ensemble average. For the stationary, isotropic, unpolarized, and Gaussian GW background, the cross-correlation can be computed analytically~\cite{Hellings:1983fr}. The result is
\begin{align}\label{eq:SGWB}
	C^{\rm SGWB}_{ab}(\tau) &=\sum_{i} \Gamma_{\rm HD}(\xi) \Phi_{\rm GW}(f_i)\cos2\pi f_i\tau~,
\end{align}
with
\begin{align}\label{eq:PhiSGWB}
     \Phi_{\rm GW}(f_i) &= \frac{1}{12\pi^2 f_i^3}\frac{1}{T_{\rm obs}}h_{c}^2(f_i)~.
\end{align}
Note that we discretized the frequency integral into the sum of the frequency bins determined by the observation time $T_{\rm obs}$, as in~\cite{NANOGrav:2023gor}.
Here, $\Gamma_{\rm HD}$ is the Hellings-Downs curve given by
\begin{align}
    \Gamma_{\rm HD}(\xi) &= \frac{1}{2} + \frac{3}{2}\left(\frac{1-\cos\xi}{2}\right)\log\left(\frac{1-\cos\xi}{2}\right) - \frac{1}{4}\left(\frac{1-\cos\xi}{2}\right)~
\end{align}
with $\xi$ being the angular separation between pulsar $a$ and $b$, and $h_c$ is the characteristic strain amplitude of the stochastic GW background defined by
\begin{align}
	\braket{h_{P}(f,\bm{\Omega})h_{P'}^*(f',\bm{\Omega}')} &= \frac{h_{c}^2(f)}{16\pi f} \delta(f-f')\delta_{PP'}\delta(\bm{\Omega} - \bm{\Omega}')~.
\end{align}
The characteristic strain is related to the energy density spectrum normalized by the critical density of the universe $\Omega_{\rm GW}(f)$ as~\cite{Allen:1997ad}
\begin{align}
	\Omega_{\rm GW}(f) &= \frac{2 \pi^2}{3 H_0^2}f^2 h_c^2(f)~,
\end{align}
where $H_0$ denotes the Hubble constant.
It is conventional to parametrize the characteristic strain as 
\begin{align}\label{eq:hcpower}
    h_c(f) = A_{\rm GW} \bigg( \frac{f}{f_{\rm ref}} \bigg)^\alpha 
\end{align}
with a reference frequency $f_{\rm ref}$. 
Then, Eq.~\eqref{eq:PhiSGWB} is rewritten as
\begin{align}\label{eq:PhiGW2}
     \Phi_{\rm GW}(f_i) &= \frac{A_{\rm GW}^2}{12\pi^2 }\frac{1}{T_{\rm obs}} 
     \bigg( \frac{f_i}{f_{\rm ref}} \bigg)^{-\gamma} f_{\rm ref}^{-3}~, 
\end{align}
where $\gamma = -2\alpha + 3$.

Now let us proceed to calculate the cross-correlation signal of the vector dark matter. When taking the time average, one should note that $\mu^{-1} \lesssim 10~{\rm yr}$ for the dark matter with the mass $\mu \gtrsim 10^{-24}~{\rm eV}$. This means that after the time average over the observational period ($T_{\rm obs} \sim 10~{\rm yr}$), oscillating terms such as $\cos(2\mu t)$ vanish. With this in mind, we substitute Eq.~\eqref{eq:residual2} to Eq.~\eqref{eq:defcorr} and take time average to obtain
\begin{align}
    \braket{\Delta T_a(t)\Delta T_b(t+\tau)} &= \frac{1}{2(2\mu)^2}\left[ \braket{\Psi_{\rm osc}(\bm{x}_{a})\Psi_{\rm osc}(\bm{x}_{b})}\braket{F_{a,{\rm P}}^{\rm DM}F_{b,{\rm P}}^{\rm DM}}\right.\cr
    &\qquad\qquad\qquad \left. \times \cos \left(2\mu \tau +2\mu(L_a -L_b)- 2\bm{k}\cdot (\bm{x}_a - \bm{x}_b)\right) \right.\cr
    &\left. \qquad\qquad  - \braket{\Psi_{\rm osc}(\bm{x}_{\rm E})\Psi_{\rm osc}(\bm{x}_b)}\braket{F_{a,{\rm E}}^{\rm DM}F_{b,{\rm P}}^{\rm DM}}
    \right. \cr
    &\left. \qquad\qquad\qquad\quad
    \times \cos(2\mu \tau -2\mu L_b- 2 \bm{k}\cdot (\bm{x}_{\rm E} - \bm{x}_b))\right.\cr
    &\left. \qquad\qquad\quad - \braket{\Psi_{\rm osc}(\bm{x}_{\rm E})\Psi_{\rm osc}(\bm{x}_a)}\braket{F_{a,{\rm P}}^{\rm DM}F_{b,{\rm E}}^{\rm DM}}
    \right.\cr 
    &\left. \qquad\qquad\qquad\qquad 
    \times \cos(2\mu \tau+2\mu L_a - 2 \bm{k}\cdot (\bm{x}_a - \bm{x}_{\rm E}))\right.\cr
    &\left. \qquad\qquad\qquad + \braket{\Psi_{\rm osc}^2(\bm{x}_{\rm E})}\braket{F_{a,{\rm E}}^{\rm DM}F_{b,{\rm E}}^{\rm DM}}\cos 2\mu \tau\right]\cr
    &\approx \frac{1}{2(2\mu)^2} \braket{\Psi_{\rm osc}^2(\bm{x}_{\rm E})}\braket{F_{a,{\rm E}}^{\rm DM}F_{b,{\rm E}}^{\rm DM}}\cos 2\mu \tau~.
\end{align}
Here, terms other than the correlation between the Earth terms vanish by averaging over the distance to the pulsars. 

The pulsars we observe are distributed all over the sky; thus, their relative angle to the direction of the vector field is completely random. This means that the average of the cross-correlation signals among many pulsars effectively takes an average over the direction of the vector field $\bm{\Omega}_A$. To calculate the angular dependence of the cross-correlation signal, we choose the coordinate system such that $\bm{u}_a = (0,0,1)$ and $\bm{u}_b = (\sin\xi,0,\cos\xi)$. In this coordinate system, we can easily perform the integral over $\bm{\Omega}_A$ to obtain
\begin{align}
    \braket{F_{a,{\rm E}}^{\rm DM}F_{b,{\rm E}}^{\rm DM}} &= 9\int \frac{d\bm{\Omega}_A}{4\pi}\ \left(1 - 4 (\bm{u}_a \cdot \bm{\Omega}_{A, {\rm E}})^2 \right)\left(1 - 4 (\bm{u}_b \cdot \bm{\Omega}_{A, {\rm E}})^2 \right)\cr
    &= \frac{3}{5}\left(7 + 16 \cos2\xi\right)\equiv \frac{138}{5}\Gamma_{\rm DM}(\xi)~.
\end{align}
Here, 
\begin{align}\label{eq:ACDM}
    \Gamma_{\rm DM}(\xi) &= \frac{5}{138}P_0(\cos\xi) + \frac{64}{138}P_2(\cos\xi)
\end{align}
shows the angular correlation between the pulsars, and we normalized it such that $\Gamma_{\rm DM}(0) = 1/2$. Eq.~\eqref{eq:ACDM} shows that the vector field will introduce a mixture of the monopole and the quadrupole angular correlation pattern.

\begin{figure}
\centering
\includegraphics[keepaspectratio, scale=0.7]{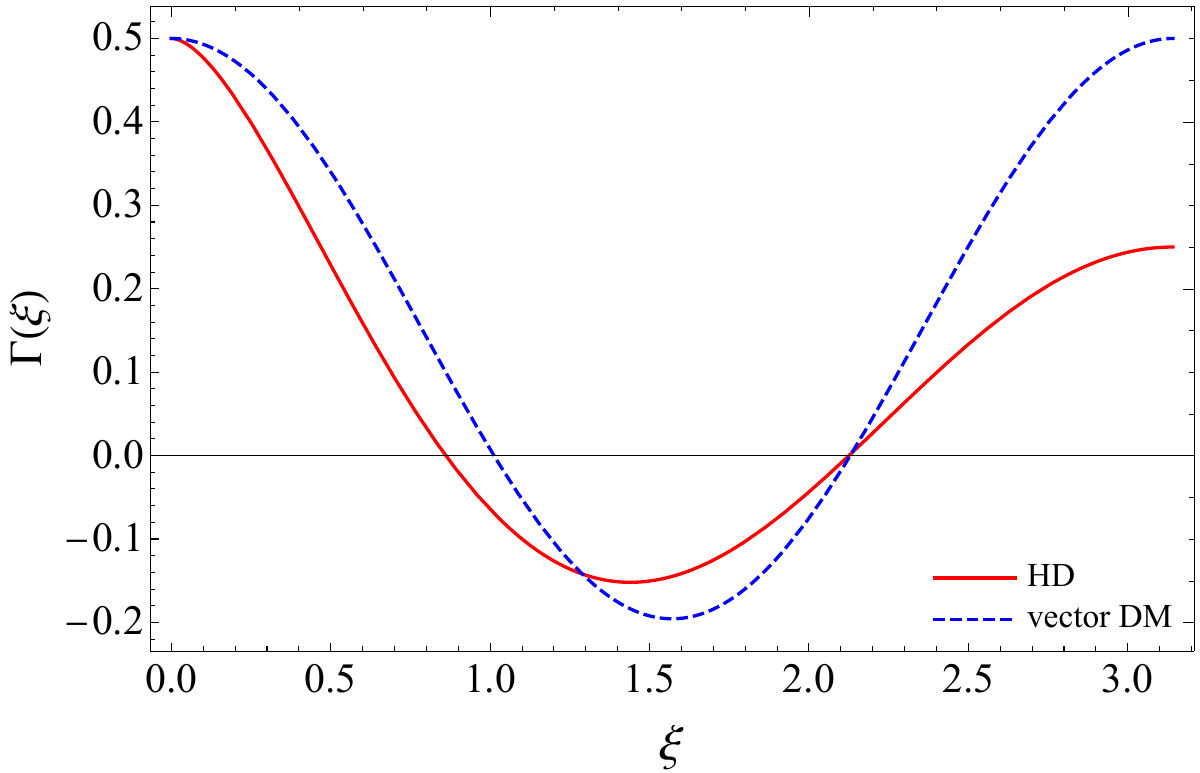}
\caption{The red solid and blue dashed curves correspond to the angular correlation of the timing residuals induced by the gravitational wave background (Hellings-Downs curve) and the ultralight vector dark matter, respectively. } 
\label{fig:1}
\end{figure}

To summarize, the cross-correlation signal produced by the ultralight vector field is given by
\begin{align}
	C^{\rm DM}_{ab}(\tau) &= \Gamma_{\rm DM}(\xi)\Phi_{\rm DM} \cos 2\mu \tau~,
\end{align}
where
\begin{align}
   \Phi_{\rm DM} = \frac{69}{5(2\mu)^2} \braket{\Psi_{\rm osc}^2(\bm{x}_{\rm E})}~.
\end{align}
In Fig.~\ref{fig:1}, the Hellings-Downs pattern due to stochastic GWs, $\Gamma_{\rm HD}(\xi)$, is drawn by a red solid curve, and the angular correlation pattern produced by the vector dark matter, $\Gamma_{\rm DM} (\xi)$, is drawn by a blue dashed curve.
We observe the angular correlation produced by the vector dark matter is dominated by the quadrupole, which is evident from the expression~\eqref{eq:ACDM} (see also Fig.~\ref{fig:ang}).

The total correlation signal is given by the sum of the signal from GWs and the ultralight vector dark matter, which is
\begin{align}\label{eq:corrf}
	C_{ab}(\tau) = \sum_i \Gamma_{\rm HD}(\xi) \Phi_{\rm GW}(f_i)\cos 2\pi f_i \tau
 + \Gamma_{\rm DM}(\xi) \Phi_{\rm DM} \cos 2\mu \tau~.
\end{align}
Now, we define the effective overlap reduction function $\Gamma_{\rm eff}$ at the frequency band which includes $2\pi f = 2\mu$ as
\begin{align}
	 \Gamma_{\rm eff}(\xi) = \frac{\Phi_{\rm GW}(\mu/\pi)}{\Phi_{\rm GW}(\mu/\pi) + \Phi_{\rm DM}}
	  \left(\Gamma_{\rm HD}(\xi) + \frac{\Phi_{\rm DM}}{\Phi_{\rm GW}(\mu/\pi)} \Gamma_{\rm DM}(\xi)\right)~.
\end{align}
The normalization is chosen so that $\Gamma_{\rm eff}(0) = 1/2$.

\begin{figure}
\centering
\includegraphics[keepaspectratio, scale=0.7]{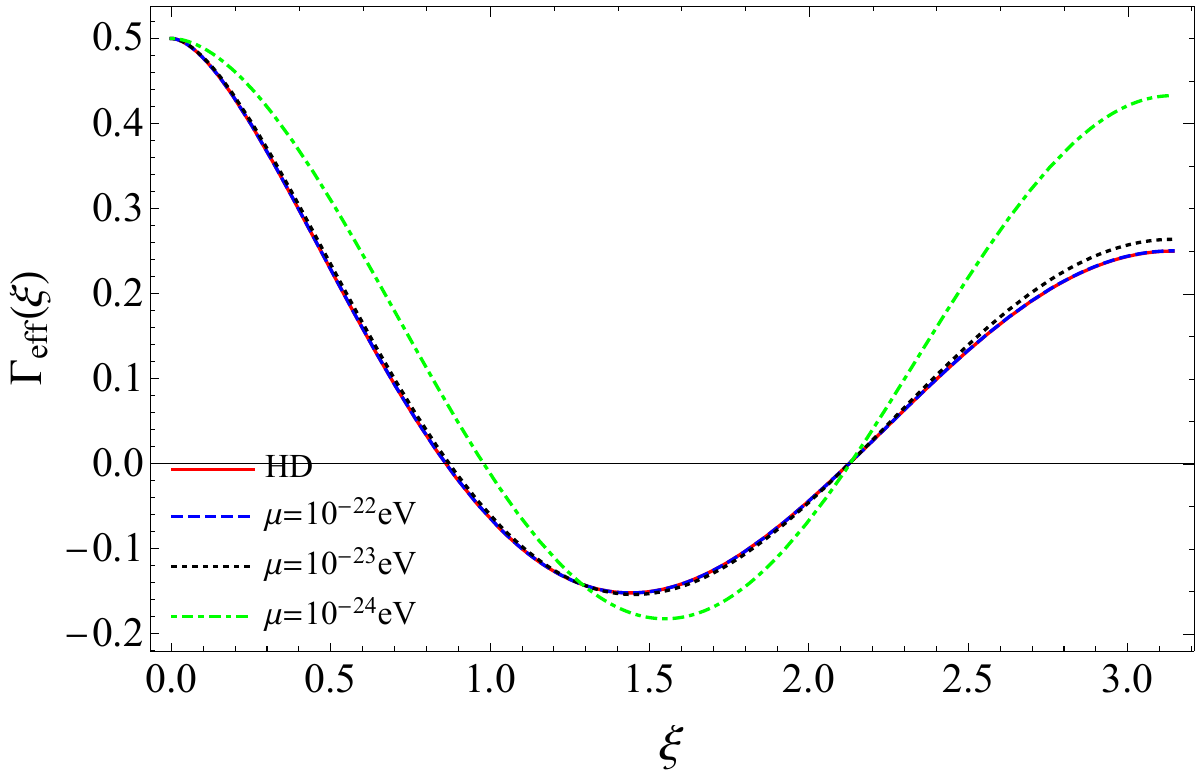}
\caption{The blue dashed, the black dotted, and the green dot-dashed curves correspond to $\Gamma_{\rm eff}(\xi)$ for $\mu = 10^{-22}, 10^{-23},$ and $10^{-24}~{\rm eV}$, respectively. 
Here, we take $\rho_{\rm DM} = 0.4 ~ {\rm GeV} \cdot {\rm cm}^{-3}$ and $T_{\rm obs} = 15 ~ {\rm yr}$.
For reference, we show the Hellings-Downs curve in the red solid curve.} 
\label{fig:2}
\end{figure}

For $A_{\rm GW}$ introduced in Eq.~\eqref{eq:hcpower}, the NANOGrav obeservation~\cite{NANOGrav:2023gor} reported $A_{\rm GW} = 2.4 \times 10^{-15}$ at $\gamma = 13/3$ with $T_{\rm obs} \sim 15 ~{\rm yr}$ and $f_{\rm ref} = 1 ~{\rm yr}^{-1}$.
Using this value, through Eq.~\eqref{eq:PhiGW2}, $\Phi_{\rm GW}$ at $f = \mu/\pi$ is estimated to be
\begin{align}
	\Phi_{\rm GW}(\mu/\pi) &\sim 5 \times 10^{-34}{\rm yr}^{2}\left(\frac{\mu}{10^{-22} {\rm eV}}\right)^{-13/3}\left(\frac{15 {\rm yr}}{T_{\rm obs}}\right)~.
\end{align}
On the other hand, assuming the dark matter is composed solely of the ultralight dark matter and neglecting the stochastic effect~\cite{Centers:2019dyn}, we can estimate $\Phi_{\rm DM}$ as 
\begin{align}
	\Phi_{\rm DM} &\sim  7 \times 10^{-37}{\rm yr}^2 \left(\frac{\rho_{\rm DM}}{0.4 {\rm GeV \cdot cm^{-3}}}\right)^2 \left(\frac{10^{-22} {\rm eV} }{\mu}\right)^6~.
\end{align}
In Fig.~\ref{fig:2}, We show $\Gamma_{\rm eff}$ for several values of the ultralight dark matter mass $\mu$. Our result shows that the presence of the ultralight vector field will deform the angular correlation pattern from the Helling-Downs curve if the vector field mass is as light as $\mu \sim 10^{-23}{\rm eV}$. However, for the heavier vector field mass, the contribution from the dark matter is negligible, and thus no characteristic feature from the dark matter can be easily observed.

\section{Conclusion}
The Hellings-Downs correlation plays a central role in confirming the existence of stochastic gravitational waves (GWs). Hence, it is important to clarify possible contamination from other effects. 
Given that pulsar timing arrays (PTAs) are sensitive both to GWs and ultralight dark matter at nanohertz frequencies, it is necessary to examine the effect of ultralight dark matter on PTA angular correlations.
For this reason, we studied the Hellings-Downs angular correlation in the presence of GWs and ultralight vector dark matter.

Our main finding is that a deformation of the Hellings-Downs correlation curve can be induced by ultralight vector dark matter at the frequency $f = \mu/\pi$ if the dark matter mass is in the range $10^{-24}~{\rm eV} \lesssim \mu \lesssim 10^{-23}~{\rm eV}$. This deformation can be a clear signature of the ultralight vector dark matter. For the higher dark matter mass, the contamination due to the dark matter is small enough; thus, the Hellings-Downs curve is maintained.

This paper focused on the angular correlation pattern produced by the vector dark matter. Let us now discuss possible applications of our findings. The immediate application is to constrain the vector dark matter by the angular correlation of PTAs. In the previous literature, the angular correlation induced by ultralight dark matter has been ignored. Since the angular correlation would provide more information, the constraint should be strengthened. Also, it is possible to consider the prospect of separately observing the GW background and the dark matter in a similar way to the decomposition of polarization components in the GW background~\cite {Kato:2015bye, Cornish:2017oic, Belgacem:2020nda}.

Although we have concentrated on the spin-1 matter, extending the analysis to higher spin matters is straightforward.
For example, there have been some attempts to consider the spin-2 matter as dark matter~\cite{Aoki:2016zgp, Aoki:2017cnz, Babichev:2016bxi, Marzola:2017lbt, Manita:2022tkl}. Moreover, the pulsar timing residual has been calculated with higher spin fields~\cite{Sun:2021yra, Unal:2022ooa}. The higher spin fields should produce different angular correlation patterns, which can be used to constrain them. 

Another simple extension of our work is to consider anisotropic components of the signal. Since the vector dark matter naturally introduces the specific direction, there should be a strong anisotropy in pulsar timing signals. We leave these issues for future work.

\begin{acknowledgments}
H.\ O. was supported by JSPS KAKENHI Grant Number JP22J14159.
K.\ N. was supported by JSPS KAKENHI Grant Number JP21J20600.
J.\ S. was in part supported by JSPS KAKENHI Grant Numbers JP17H02894, JP17K18778, JP20H01902, JP22H01220.
\end{acknowledgments}

\appendix*

\bibliography{ref}

\end{document}